\documentclass[twocolumn,showpacs,amsmath,amssymb,floatfix,aps,pra]{revtex4}
\usepackage{graphicx}
\def\ep{\varepsilon}
\begin{document}

\title{Fault-Tolerant Landau-Zener Quantum Gates}

\author{C.~Hicke, L.~F.~Santos\footnote{Present address: Department
of Physics and Astronomy, Dartmouth College, Hanover, NH 03755}, and
M.~I.~Dykman} \affiliation{Department of Physics and Astronomy,
Michigan State University, East Lansing, MI 48824}

\date{\today}

\begin{abstract}
We present a method to perform fault-tolerant single-qubit gate
operations using Landau-Zener tunneling. In a single Landau-Zener
pulse, the qubit transition frequency is varied in time so that it
passes through the frequency of the radiation field. We show that a
simple three-pulse sequence allows eliminating errors in the gate up
to the third order in errors in the qubit energies or the radiation
frequency.
\end{abstract}
\pacs{03.67.Lx, 82.56.Jn, 03.67.Mn} \maketitle

\section{Introduction}

In many proposed implementations of a quantum computer (QC)
single-qubit operations are performed by applying pulses of
radiation. The pulses cause resonant transitions between qubit
states, that is between the states of the system that comprises a
qubit. The operation is determined by the pulse amplitude and
duration. In many proposals, particularly in the proposed scalable
condensed-matter based systems \cite{Kane1998}, control pulses will
be applied globally, to many qubits at a time. A target qubit is
chosen by tuning it in resonance with the radiation. The
corresponding gate operations invariably involve errors which come
from the underlying errors in the frequency, amplitude, and length
of the radiation pulse as well as in the qubit tuning.

Improving the accuracy of quantum gates and reducing their
sensitivity to errors from different sources is critical for a
successful operation of a QC. Much progress has been made recently
in this direction by using radiation pulses of special shape and
composite radiation pulses \cite{Vandersypen2004}. In the analysis
or resonant pulse shape it is usually assumed that the qubit
transition frequency is held constant during the pulse.

An alternative approach to single-qubit operations is based on
Landau-Zener tunneling (LZT) \cite{Landau1932,Zener1932}. In this
approach the qubit transition frequency $\omega_0(t)$ is swept
through the frequency of the resonant field $\omega_F$
\cite{Dykman2001b}. The change of the qubit state depends on the
field strength and the speed at which $\omega_0(t)$ is changed when
it goes through resonance \cite{Benza2003}. The LZT can be used also
for a two-qubit operation in which qubit frequencies are swept past
each other leading to excitation swap
\cite{Dykman2001b,Platzman1999,Saito2004}.

In the present paper we study the robustness of the LZT-based gate
operations. We develop a simple pulse sequence that is extremely
stable against errors in the qubit transition frequency or
equivalently, the radiation frequency. Such errors come from various
sources. An example is provided by systems where the qubit-qubit
interaction is not turned off, and therefore the transition energy
of a qubit depends on the state of other qubits. Much effort has
been put into developing means for correcting them using active
control \cite{Viola2004,Facchi2005,Sengupta2005}.

An advantageous feature of LZT is that the change of the qubit state
populations depends on the radiation amplitude and the speed of the
transition frequency change $\dot\omega_0$, but not on the exact
instant of time when the frequency coincides with the radiation
frequency, $\omega_0(t)=\omega_F$. However, the change of the phase
difference between the states depends on this time. Therefore an
error in $\omega_0$ or $\omega_F$ leads to an error in the phase
difference, i.e., a phase error. This error has two parts: one comes
from the phase accumulation before crossing the resonant frequency,
and the other after the crossing. Clearly, they have opposite signs.

A natural way of reducing a phase error is to make the system
accumulate the appropriate opposite in sign phases before and after
the ``working" pulse. To do this, we first apply a strong radiation
pulse that swaps the states, which can be done with exponentially
high efficiency using LZT. Then we apply the ``working" pulse, and
then another swapping pulse. The swapping pulses effectively change
the sign of the accumulated phase. As we show, by adjusting their
parameters we can compensate phase errors with a high precision.

In Sec.~II below we give the scattering matrix for LZT in a modified
adiabatic basis which turns out to be advantageous compared to the
computational basis. The scattering matrix describes the quantum
gate. In Sec.~III it is presented in more conventional for quantum
computation terms of the qubit rotation matrix. In Sec.~IV, which is
the central part of the paper, we propose a simple composite
Landau-Zener (LZ) pulse and demonstrate that it efficiently compensates
energy offset errors even where these errors are not small. Sec.~V
contains concluding remarks.

\section{Landau-Zener transformation in the modified adiabatic basis}

A simple implementation of the LZ gate is as follows. The
amplitude of the radiation pulse is held fixed, while the difference
between the qubit transition frequency and the radiation frequency
\begin{equation}
\label{eq:freq_definition} \Delta =\Delta (t)=\omega_F-\omega_0(t)
\end{equation}
is swept through zero.  If $\omega_0(t)$ is varied slowly compared
to $\omega_F$, i.e., $|\dot\omega_0|\ll \omega_F^2$, the qubit
dynamics can be described in the rotating wave approximation, with
Hamiltonian
\begin{equation}
\label{eq:Hamiltonian} H = H(t)=\left(
\begin{array}{cc}
   \Delta /2 & \gamma   \\
   \gamma  &  - \Delta /2
\end{array} \right).
\end{equation}
Here, $\gamma$ is the matrix element of the radiation-induced
interstate transition. The Hamiltonian $H$ is written in the
so-called computational basis, with wave functions $\left| 0
\right\rangle = \left(
\begin{array}{c}    1  \\    0 \end{array}\right)$ and
$\left| 1 \right\rangle = \left( \begin{array}{c}
   0  \\    1 \end{array} \right) $.

We assume that well before and after the frequency crossing the
values of $|\Delta|$ largely exceed $\gamma$ and $\Delta$ slowly
varies in time, $|\dot\Delta/\Delta^2|\ll 1$. Then the wave
functions of the system are well described by the adiabatic
approximation, i.e., by the instantaneous eigenfunctions of the
Hamiltonian (\ref{eq:Hamiltonian}),
\begin{eqnarray}
\label{eq:basis} &&\left| {\psi _{0} } \right\rangle  = \left[
\begin{array}{c}
   \cos(\theta/2)  \\
\sin(\theta/2)
\end{array} \right], \qquad
\left| {\psi _{1} } \right\rangle  = \left[
\begin{array}{c}
   -\sin(\theta/2)  \\
\cos(\theta/2)
\end{array} \right],\\
&&\theta=({\rm sgn}\Delta)\cos^{-1}\frac{|\Delta|}{2E}, \qquad
E=\left(\frac{\Delta^2}{4}+\gamma^2\right)^{1/2},\nonumber
\end{eqnarray}
where $\Delta\equiv \Delta(t)$ and $(-1)^n E\,{\rm sgn}\Delta$ is
the adiabatic energy of the states
$|\psi_n\rangle=|\psi_{0,1}\rangle$. The adiabatic approximation for
$E$ and $\theta$ is accurate to $\gamma^2\dot\Delta/\Delta^3$ and
$\gamma\dot\Delta/\Delta^3$, respectively.

In contrast to the standard adiabatic approximation, we chose the
states $|\psi_{0,1}\rangle$ and their energies in such a way that
$|\psi_0\rangle$ and $|\psi_1\rangle$ go over into $|0\rangle$ and
$|1\rangle$, respectively, for $|\Delta|/\gamma\to\infty$. As a
result $\theta$ is discontinuous as a function of $\Delta$ for
$\Delta=0$, but the adiabatic approximation does not apply for such
$\Delta$ anyway.

For the future analysis it is convenient to introduce the Pauli
matrices $X,Y,Z$ in the basis (\ref{eq:basis}), with
\[Z|\psi_n\rangle= (1-2n)|\psi_n\rangle, \qquad X|\psi_n\rangle=
|\psi_{1-n}\rangle \quad(n=0,1),\]
and $Y=iXZ$. In these notations, the operator of the adiabatic time
evolution $U(t_f,t_i)=T\exp[-i\int\nolimits_{t_i}^{t_f} dt H(t)]$
has the form
\begin{eqnarray}
\label{eq:propagator} U(t_f,t_i)=\exp\left[-i\left({\rm
sgn}\,\Delta\right)Z\,\int_{t_i}^{t_f}E(t)\,dt\right],
\end{eqnarray}
with  ${\rm sgn}\,\Delta\equiv {\rm sgn}\,\Delta(t_i) \equiv {\rm
sgn} \,\Delta(t_f)$ [the sign of $\Delta(t)$ is not changed in the
range where Eq.~(\ref{eq:propagator}) applies].

The LZ transition can be thought of as occurring between
the states (\ref{eq:basis}). Following the standard scheme
\cite{Landau1932,Zener1932} we take two values $\Delta_{1,2}$ of
$\Delta(t)$ such that they have opposite signs, $\Delta_1\Delta_2 <
0$. We choose $|\Delta_{1,2}|$ sufficiently large, so that the
adiabatic approximation (\ref{eq:basis}) applies for
$\Delta(t_i)=\Delta_i, i=1,2$. At the same time, $|\Delta_{1,2}|$
are sufficiently small, so that $\Delta(t)$ can be assumed to be a
linear function of time between $\Delta_1$ and $\Delta_2$,
\begin{equation}
\label{eq:freq_difference} \Delta(t)\approx -\eta (t-t_c), \qquad
\eta = -\dot\Delta(t_c),
\end{equation}
where the crossing time $t_c$ is given by the condition
$\Delta(t_c)=0$. The adiabaticity for $t=t_{1,2}$ requires that
$|\Delta_{1,2}|\gg \gamma,\eta^{1/2}$. We will consider the
LZ transition first for the case $\Delta_1>0$ and
$\Delta_2<0$, when $\eta>0$.

\begin{figure}[h]
\includegraphics[width=3.2in]{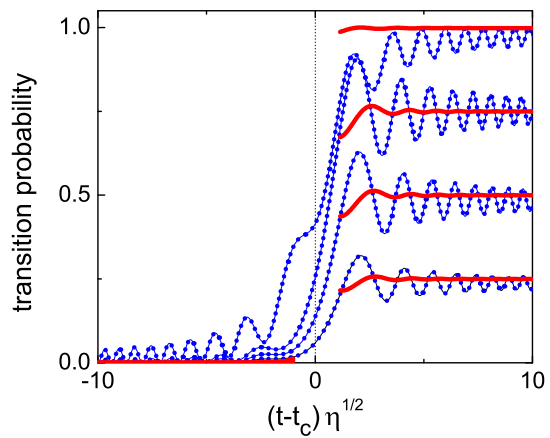}
\caption{(Color online). Landau-Zener transitions
$|\psi_0\rangle\to|\psi_1\rangle$ and $|0\rangle\to|1\rangle$ in the
modified adiabatic basis (\protect\ref{eq:basis}) and in the
computational basis for linear $\Delta(t)$
(\ref{eq:freq_difference}). Solid and dotted lines show the squared
amplitude of the initially empty states $|\psi_1\rangle$ and
$|1\rangle$, respectively. The data for the state $|\psi_1\rangle$
close to $\Delta(t)=0$ are not shown, since the adiabatic
approximation does not apply for small $|\Delta|$. The lines refer to
$g=1, 0.47, 0.33,$ and $0.21$, in the order of decreasing transition
probability for $(t-t_c)\eta^{1/2}=10$. As long as $\Delta(t)$ is
large and negative, the system stays in the initially occupied
adiabatic state $|\psi_0\rangle$, and therefore the solid curves for
different $g$ cannot be resolved for
$(t-t_c)\eta^{1/2}\lesssim-1$. For large $(t-t_c)\eta^{1/2}$ the solid
lines quickly approach the Landau-Zener probabilities $1-e^{-2\pi
g^2}$.} \label{fig:crossing}
\end{figure}

The modified adiabatic basis (\ref{eq:basis}) is advantageous,
because in this basis the transition matrix $S$ has a particularly
simple form. For $\Delta(t)$ of the form (\ref{eq:freq_difference})
the error in $S$ is determined by the accuracy of the adiabatic
approximation itself and is of order $\gamma/\eta^2|t_{1,2}-t_c|^3$,
in contrast to the computational basis, where the error is $\sim
O(\gamma/\eta|t_{1,2}-t_c|)$. This latter error is comparatively
large for the values of $\gamma/|\Delta_{1,2}|$ of interest for
quantum computing. It leads to the well-known oscillations of the transition amplitude with increasing $|\Delta|$
\cite{Benza2003}, whereas in the basis (\ref{eq:basis}) such
oscillations do not arise, see Fig.\ref{fig:crossing}.

The energy detuning $|\Delta_{1,2}|$ cannot be made too large,
because this would make the gate operation long. If we characterize
the overall error of the adiabatic approximation as the sum
$\sum\nolimits_{i=1,2}\gamma/\eta^2|t_i-t_c|^3$ and impose the
condition that the overall duration of the operation $t_2-t_1$ be
minimal, we see that the error is minimized when the pulses
$\Delta(t)$ are symmetrical, $t_2-t_c=t_c-t_1$, i.e.,
$|\Delta_1|=|\Delta_2|$.

The matrix $S(t_2,t_1)\equiv S$ in the basis (\ref{eq:basis}) can be
obtained using the parabolic cylinder functions that solve the
Schr\"odinger equation with the Hamiltonian (\ref{eq:Hamiltonian}),
(\ref{eq:freq_difference}),
\begin{widetext}
\begin{equation}
\label{eq:scattering_matrix} S(t_2,t_1) =\left(
\begin{array}{cc}
   \exp\left[ - \pi g^2  + i(\varphi_2  - \varphi_1 )\right] &  -
   \frac{(2\pi)^{1/2}    }{g \,\Gamma \left( ig^2\right)}
\exp\left[ - \frac{\pi}{2} g^2- i\frac{\pi}{4}+i(\varphi_1+\varphi_2)\right]    \\
   \frac{(2\pi)^{1/2}
   }
{g \,\Gamma \left( -ig^2\right)} \exp\left[ - \frac{\pi}{2} g^2+
i\frac{\pi}{4}- i(\varphi_1+\varphi_2)\right]   & \exp\left[ - \pi
g^2  - i(\varphi_2 - \varphi_1 )\right]
\end{array}\right),
\end{equation}
\end{widetext}
where $\Gamma(x)$ is the gamma function.

The dimensionless coupling parameter $g  = \gamma/|\eta|^{1/2}$ in
Eq.~(\ref{eq:scattering_matrix}) is the major parameter of the
theory, it determines the amplitude of the $|\psi_n\rangle \to
|\psi_{1-n}\rangle$ transition. The phases $\varphi_{1,2}$ are
\begin{equation}
\label{eq:phases} \varphi_i  = \frac{\Delta_i^2 } {4|\eta|} + g^2\ln
\left(\frac{|\Delta_i|}{|\eta|^{1/2}}
\right)+\frac{g^4|\eta|}{2\Delta_i^2}, \qquad i = 1,2.
\end{equation}
Here we have disregarded the higher order terms in
$|\Delta_{1,2}|^{-1}$. The constants in $\varphi_{1,2}$ are chosen
so as to match the corresponding constants in the parabolic cylinder
functions \cite{Gradshteyn2000}.

The matrix $S$ for a transition from the initial state with
$\Delta_1<0$ to the final state with $\Delta_2>0$ is given by the
transposed matrix~(\ref{eq:scattering_matrix}) in which the phases
$\varphi_1$ and $\varphi_2$ are interchanged. In this case $\eta <
0$ in Eq.~(\ref{eq:freq_difference}); the expressions for
$\varphi_{1,2}$ and $g$ do not change.

\section{Rotation matrix representation}

The LZ transition can be conveniently described using the
standard language of gate operations in quantum computing. To do
this we express the transition matrix in terms of the operators
$R_{X}(\theta)=\exp(-i\theta X/2)$ and $R_{Z}(\theta)=\exp(-i\theta
Z/2)$ of rotation about $x$ and $z$ axes in the basis
(\ref{eq:basis}). The rotation matrices can be written using the
``adiabatic'' phases $\phi(t_i)$ that accumulate between the time
$t_i$ and the time $t_c$ at which the levels would cross in the
absence of coupling. From  Eq.~(\ref{eq:phases})
\begin{eqnarray}
\label{eq:phases_with_t_c} &&\varphi_i= \phi(t_i) +\varphi_0 \quad
(i=1,2),
\qquad t_1<t_c<t_2,\nonumber\\
&&\phi(t_i)= \left\vert\int_{t_c}^{t_i}E\,dt\right\vert,
\qquad\varphi_0=\frac{1}{2}g^2(\ln g^2-1),
\end{eqnarray}
where we have disregarded corrections $\propto|\Delta_{1,2}|^{-4}$,
in agreement with the approximations made in obtaining
Eq.~(\ref{eq:scattering_matrix}).

For the case $\Delta_1>0> \Delta_2$ the dependence of the transition
matrix $S$ (\ref{eq:scattering_matrix}) on the phases
$\phi(t_{1,2})$ has the form
\begin{equation}
\label{eq:S_to_S_prime}
S(t_2,t_1)=R_z\left[-2\phi(t_2)\right]S^{\prime}R_z\left[2\phi(t_1)\right].
\end{equation}
A direct calculation shows that the matrix $S^{\prime}$ is
\begin{equation}
\label{eq:S_prime} S^{\prime}=R_z(\Phi)R_x(\alpha)R_z(-\Phi).
\end{equation}
The rotation angles $\Phi,\alpha$ are given by the expressions
\begin{eqnarray}
\label{eq:angles} &&\Phi=-2\varphi_0 + \arg \Gamma(ig^2) +
\frac{3\pi}{4},\\
&&\alpha = 2\cos^{-1}\left[\exp(-\pi g^2)\right].\nonumber
\end{eqnarray}

A minor modification of these equations allows using them also for
the case $\Delta_1<0<\Delta_2$ when the frequency difference is
increased in time in order to bring the states in resonance. It was
explained below Eq.~(\ref{eq:phases}) how to relate the matrix $S$
in this case to the matrix $S$ for $\Delta_1 > 0 >\Delta_2$.
Following this prescription we obtain
\begin{equation}
\label{eq:negative_S}
S(t_2,t_1)=R_z\left[2\phi(t_2)-\Phi\right]R_x(\alpha)
R_z\left[-2\phi(t_1)+\Phi\right].
\end{equation}
In the rotation matrix representation, the only difference from the
$S$ matrix from the case of decreasing $\Delta(t)$ is that $\Phi$
and $\phi(t_{1,2})$ change signs.
\begin{figure}[h]
\includegraphics[width=3.2in]{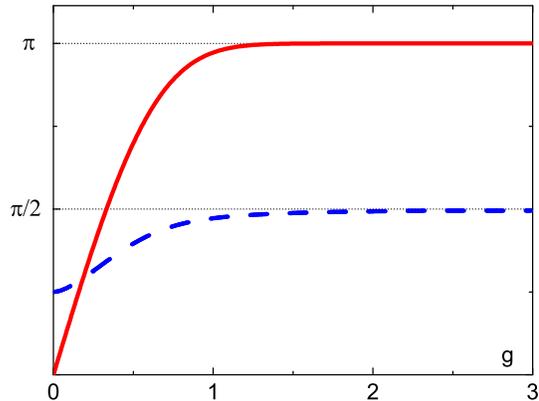}
\caption{(Color online). The rotation angles $\alpha$ (solid line) and $\Phi$
(dashed line) in the rotation-matrix representation of the
Landau-Zener gate operation as functions of the control parameter
$g$. The $\pi/2$ gate, $\alpha=\pi/2$, requires $g=[\ln
2/(2\pi)]^{-1/2}\approx 0.33$.} \label{fig:fig1_LZ}
\end{figure}
Eqs.~(\ref{eq:S_to_S_prime})-(\ref{eq:negative_S}) express the
LZ transition matrix in the form of rotation operators in
the basis of the modified adiabatic states $|\psi_0\rangle$ and
$|\psi_1\rangle$ (\ref{eq:basis}). For strong coupling, $\exp(-\pi
g^2)\ll 1$, the rotation angle $\alpha$ approaches $\pi$, which
corresponds to a population swap between the adiabatic states. It is
well known from the LZ theory \cite{Landau1932,Zener1932}
that the swap operation is exponentially efficient, $\pi-\alpha
\approx 2\exp(-\pi g^2)$ for large $g$. In the opposite limit of
weak coupling, $g\ll 1$, the change of the state populations is
small, $\alpha \approx (8\pi)^{1/2}g$. In addition to the change of
state populations there is also a phase shift that accumulates
during an operation.  The dependence of the angles $\alpha$ and
$\Phi$ on the coupling parameter $g$ is shown in
Fig.\ref{fig:fig1_LZ}.

\section{Composite Landau-Zener pulses}

For many models of quantum computers an important source of errors
are errors in qubit transition frequencies $\omega_0$. They may be
induced by a low-frequency external noise that modulates the
interlevel distance. They may also emerge from errors in the control
of the qubit-qubit interaction: if the interaction is not fully
turned off between operations, the interlevel distance is a function
of the state of other qubits. In addition there are systems where
the interaction is not turned off at all, like in liquid state
NMR-based QC's. In all these systems it is important to be able to
perform single-qubit gate operations that would be insensitive to
the state of other qubits.

The rotation-operator representation suggests a way to develop fault
tolerant composite LZ pulses with respect to errors in the
qubit transition frequency $\omega_0$ and in the radiation frequency
$\omega_F$. We will assume that there is a constant error $\ep$ in
the frequency difference $\Delta(t)=\omega_F-\omega_0(t)$, but that
no other errors occur during the gate operation. From
Eq.~(\ref{eq:freq_difference}), the renormalization $\Delta(t) \to
\Delta(t) + \ep$ translates into the change of the adiabatic energy
$E$ and the crossing time $t_c$, with $t_c\to t_c +\ep/\eta$. As a
result the phases $\phi(t_{1,2})$ as given by
Eq.~(\ref{eq:phases_with_t_c}) are incremented by

\begin{equation}
\label{eq:2nd_order_error}
 \delta\phi(t_{i}) =\frac{E(t_i)\Delta(t_i)}{|\eta\Delta(t_i)|}
 \ep +\frac{|\Delta(t_i)|}{8|\eta| E(t_i)}\ep^2,\qquad i=1,2,
\end{equation}
to second order in $\ep$.

\subsection{Error compensation with $\pi$-pulses}

\begin{figure}[h]
\includegraphics[width=3.2in]{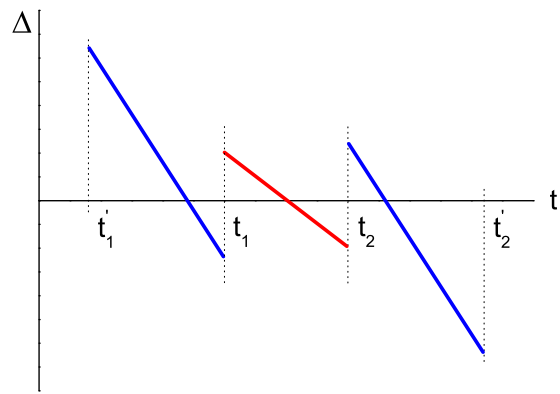}
\caption{(Color online). An idealized composite pulse. The first and third pulses
are $\pi$-pulses, the pulse in the middle performs the required gate
operation. The overall pulse compensates errors in the qubit energy
to 3rd order.} \label{fig:compositepulse}
\end{figure}

A simple and robust method of compensating errors in $\phi$ is based
on a composite pulse that consists of the desired pulse sandwiched
between two auxiliary pulses. Using $\pi$-pulses in which
$\Delta(t)$ is linear in $t$, as shown in
Fig.~\ref{fig:compositepulse}, it is possible to eliminate errors of
first and second order in $\ep$. The goal is to compensate the
factors $R_z[\pm2\delta\phi(t_{1,2})]$ in the $S$-matrix
(\ref{eq:S_to_S_prime}). We note that all other factors in $S$ are
not changed by the energy change $\ep$, which is one of the major
advantageous features of the LZ gate operation.
A $\pi$-pulse is obtained if $\exp(\pi g^2)\gg 1$, which is met
already for not too large $g$: for example, $\exp(-\pi g^2)<10^{-5}$
for $g > 1.92$.

Disregarding corrections $\sim \exp(-\pi g^2)$ we can write the
$S$-matrix for the $\pi$-pulse as
\begin{eqnarray}
\label{eq:pi}
S_{\pi}(t',t)&\approx& -iXR_z[2\phi_{\pi}(t)+2\phi_{\pi}(t')-2\Phi]\nonumber\\
&&\equiv -iR_z[-2\phi_{\pi}(t)-2\phi_{\pi}(t')+2\Phi]X,
\end{eqnarray}
where $t, t'$ are the initial and final times, and the subscript
$\pi$ indicates that the corresponding quantities refer to a
$\pi$-pulse. We assume that $\Delta(t) > 0 > \Delta(t')$.

The overall gate operation is now performed by a composite pulse
\begin{equation}
\label{eq:S_c}
S_c(t^{\prime}_2,t'_1)=S_{\pi}(t^{\prime}_2,t_2)S(t_2,t_1)
S_{\pi}(t_1,t'_1).
\end{equation}
In writing this expression we assumed that the system is switched
instantaneously between the states that correspond to the end
(beginning) of the correcting pulse and the beginning (end) of the
working pulse $S(t_2,t_1)$. The overall composite pulse is shown in
Fig.~\ref{fig:compositepulse}.
The first and the second $\pi$-pulses correct the errors
$\delta\phi$ (\ref{eq:2nd_order_error}) in the phases $\phi(t_1)$
and $\phi(t_2)$, respectively. We show how it works for $\phi(t_2)$.
From Eqs.~(\ref{eq:S_to_S_prime}), (\ref{eq:pi}), the error in
$\phi(t_2)$ will be compensated if

\[\delta\phi_{\pi}(t^{\prime}_2)+\delta\phi_{\pi}(t_2)-\delta\phi(t_2)=0.\]
To second order in $\ep$, the errors $\delta\phi$ here are given by
Eq.~(\ref{eq:2nd_order_error}) with appropriate $t_i$. The total
error will be equal to zero provided
\begin{eqnarray}
\label{eq:compensation}
\frac{E_{\pi}(t^{\prime}_2)}{\eta_{\pi}}-\frac{E(t_2)}{\eta}-
\frac{E_{\pi}(t_2)}{\eta_{\pi}}=0,\nonumber\\
\frac{|\Delta_{\pi}(t^{\prime}_2)|}{\eta_{\pi}E_{\pi}(t^{\prime}_2)}-
\frac{|\Delta(t_2)|}{\eta E(t_2)}+
\frac{\Delta_{\pi}(t_2)}{\eta_{\pi}E_{\pi}(t_2)}=0.
\end{eqnarray}

Equations (\ref{eq:compensation}) are simplified if we keep only the
lowest order terms with respect to $\gamma^2/\Delta^2$, in which
case $E(t_i)\approx |\Delta(t_i)|/2$ both for the working and the
correcting pulse. This gives
\begin{equation}
\label{eq:comp_simplified} \eta_{\pi}=2\eta,\qquad
|\Delta_{\pi}(t^{\prime}_2)|- 2|\Delta(t_2)|-\Delta_{\pi}(t_2)=0.
\end{equation}
An immediate consequence of Eq.~(\ref{eq:comp_simplified}) is that
the coupling constant $\gamma_{\pi}$ for the $\pi$-pulse should
exceed the value of $\gamma$ for the working pulse, because
$g_{\pi}\geq g$ and $\eta_{\pi}>\eta$. Another consequence is that
the $\pi$-pulse amplitude should exceed that of the working pulse.
If we choose $\Delta_{\pi}$ so that the error of the adiabatic
approximation in the $\pi$-pulse does not exceed that of the working
pulse, $\gamma_{\pi}\eta_{\pi}/|\Delta_{\pi}|^3\leq
\gamma\eta/|\Delta(t_2)|^3$, we obtain from
Eq.~(\ref{eq:comp_simplified}) the condition $\Delta_{\pi}(t_2)\geq
|\Delta(t_2)|2^{1/2}(g_{\pi}/g)^{1/3}$.

We note that the correcting pulse is asymmetric, with
$|\Delta_{\pi}(t^{\prime}_2)| > \Delta_{\pi}(t_2), 2|\Delta(t_2)|$,
as shown in Fig.~\ref{fig:compositepulse}. Another important comment
is that the proposed simple single pulse does not allow us to
correct errors of higher order in $\ep$. It is straightforward to
see that the equation for $\Delta_{\pi}(t_2),\Delta_{\pi}(t_2')$
that follows from the condition that the error $\sim\ep^3$ vanishes
is incompatible with Eqs.~(\ref{eq:comp_simplified}). However, the
terms $\propto (\ep/\gamma)^3$ contain a small factor
$g^2\gamma/\Delta(t_{1,2})^3\ll 1$. The higher-order terms in
$\ep/\gamma$ contain higher powers of the parameter
$\gamma/\Delta(t_{1,2})$. This is why compensating errors only up to
the second order in $\ep$ turns out efficient.

The analysis of the first correcting $\pi$-pulse, $S_\pi(t_1,t_1')$,
is similar to that given above. The amplitude of this pulse also
exceeds the amplitude of the working pulse. The duration of the
correcting pulses is close to the duration of the working pulse, for
$g\sim 1$ and $g_{\pi}\gtrsim 2$.

The pulse sequence (\ref{eq:S_c}) is written assuming that the radiation is switched off between the pulses and that the  switching between the working and correcting pulses is instantaneous. A generalization to a more realistic case of a nonzero switching time is straightforward. The time evolution between the pulses can be described by extra terms in the phases $\phi_{\pi}(t_1), \phi_{\pi}(t_2)$, leading to the appropriate modification of the equations for error compensation (\ref{eq:compensation}). The analysis can be also extended to the case where $\Delta(t)$ is a nonlinear function of time and the coupling $g$ depends on time. This extension requires numerical analysis; we have found for several types of $\Delta(t), g(t)$ that good error correction can still be achieved with a three-pulse sequence.

\subsection{Maximal error of the three-pulse sequence}

In order to demonstrate the error correction we will consider single
working pulses $S(t_2,t_1)$ with the overall phases
${\Phi-2\phi(t_{1,2})\equiv 0\,({\rm mod}\,2\pi)}$ in the absence of
errors, which we will denote by $S^{(0)}(t_2,t_1)$. Such pulses
describe transformations between the states (\ref{eq:basis}) with no
extra phase, that is pure $X$ rotations. We will also choose the
correcting $\pi$-pulses with the overall phase ${2\phi_{\pi}(t) +
2\phi(t')-2\Phi\equiv 0\,({\rm mod}\,2\pi)}$ in the absence of
errors, with $t,t'$ being $t_1',t_1$ and $t_2,t_2'$ for the first
and second pulse, respectively. Then in the absence of errors the
overall gate is either not affected by the correcting pulse or its
sign is changed.

The restriction on the phases provides extra constraints on the
parameters of the correcting pulses. First of all, it "discretizes"
the total duration of the pulses. For the correcting pulses we still
have a choice of $\Delta_{\pi}(t_2)$ and $\Delta_{\pi}(t_1)$. They
will be chosen maximally close to
$|\Delta(t_2)|2^{1/2}(g_{\pi}/g)^{1/3}$ and
$\Delta(t_1)2^{1/2}(g_{\pi}/g)^{1/3}$, respectively, in order to
minimize the error of the adiabatic approximation (\ref{eq:basis})
and to minimize the overall pulse duration.

We will characterize the gate error ${\cal E}$ by the spectral norm
of the difference of the operator $S$ in the presence of errors and
the ``ideal" gate operator $S^{(0)}$,
\begin{equation}
\label{eq:error_definition} {\cal E}=|| S-S^{(0)}||_2.
\end{equation}
Here, $||A||_2$ is the square root of the maximal eigenvalue of the
operator $A^{\dagger}A$. For uncorrected pulses $S=S(t_2,t_1)$,
whereas for corrected pulses $S=S_c(t_2',t_1')$. For simple
symmetric composite pulses described below, the overall sign of the
composite pulse is opposite to that of the original pulse in the
absence of errors. In this case we set $S=-S_c(t_2',t_1')$ in
Eq.~(\ref{eq:error_definition}).

For uncorrected pulses we have
\begin{equation}
\label{eq:error_uncorrected} {\cal
E}=2^{1/2}|1-n_{x1}n_{x2}-n_{y1}n_{y2}\cos\alpha|^{1/2},
\end{equation}
where ${\bf n}_i=(\cos[\delta\phi(t_i)], \sin[\delta\phi(t_i)])$ is
an auxiliary 2D unit vector ($i=1,2$).
Eq.~(\ref{eq:error_definition}) applies also in the case of
corrected pulses, but now we have to replace in the definition of
the ${\bf n}_1$ vector
\begin{equation}
\label{eq:substitution_corrected} \delta\phi(t_1) \rightarrow
\delta\phi(t_1)- \delta\phi_{\pi}(t_1) - \delta\phi_{\pi}(t_1').
\end{equation}
A similar replacement must be done in the definition of the vector
${\bf n}_2$.

For small phase errors $|\delta\phi(t_{1,2})|$ the function ${\cal
E}$ for uncorrected pulses is linear in the error. In particular, to
first order in $\ep$ for a symmetric pulse we have
$|\delta\phi(t_{1})|\approx |\delta\phi(t_{2})|\approx \ep
E(t_1)/\eta$, and ${\cal E} \approx 2|\ep|
E(t_1)\eta^{-1}\sin(\alpha/2)$. In contrast, by applying the same
arguments to a corrected pulse, we see that the gate error is
$\propto\ep^3$. As noted above, the terms $\propto\ep^3$ and
higher-order terms in $\ep$ contain a small factor. They become very
small already for not too small $\ep$.

\begin{figure}[h]
\includegraphics[width=3.2in]{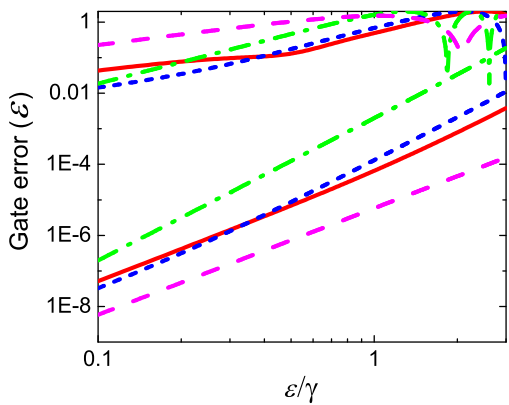}
\caption{(Color online). Gate errors ${\cal E}$ for Landau-Zener pulses as a
function of the frequency error $\ep$. The upper and lower curves
refer to the single LZ pulse and the composite pulse, respectively.
The dash-dotted, dotted, solid, and dashed lines show ${\cal E}$ for
$g= 2,1.2,1$, and $0.3$. } \label{fig:gate_error1}
\end{figure}

To illustrate how the composite pulse works we compare in
Fig.~\ref{fig:gate_error1} the error of an uncorrected LZ
gate with the gate error of the composite pulse. The data refer to
different values of $g$ of the working pulse; the corresponding
values of $\alpha$ are given in Fig.~\ref{fig:fig1_LZ}. We used
$\Delta(t_1)=-\Delta(t_2)\approx 10 \eta^{1/2}$ [the precise value
of $\Delta(t_{1,2})$ was adjusted to make $S(t_2,t_1)$ an
$X$-rotation, $S(t_2,t_1)=R_x(\alpha)$]. The compensating
$\pi$-pulses where modeled by pulses with $g_{\pi}=3$. Based on the
arguments provided at the end of Sec.~IV~A, we took
$\Delta_{\pi}(t_2)\approx |\Delta(t_2)|2^{1/2}(g_{\pi}/g)^{1/3}$,
whereas $\Delta_{\pi}(t_2')$ was found from
Eq.~(\ref{eq:compensation}); we used $\Delta_{\pi}(t_1')=
-\Delta_{\pi}(t_2')$ and $\Delta_{\pi}(t_1)=-\Delta_{\pi}(t_2)$.

It is seen from Fig.~\ref{fig:gate_error1} that the proposed
composite pulses are extremely efficient for compensating gate
errors. Even for the energy error $\ep=\gamma$, where the error of
an uncorrected pulse is close to 1, for the composite pulse ${\cal
E}\lesssim 10^{-3}$. For $g\lesssim 1$ the error of the single pulse
scales as $\ep$, whereas the error of the composite pulse scales as
$\ep^3$, in agreement with the theory. For large $g$, when the gate
is almost an $X$-gate ($\pi$-pulse), in the case of symmetric pulses
that we discuss, the coefficients at the terms $\propto\ep$ and
$\propto\ep^3$ become small; they become equal to zero for
$\alpha=\pi$. Therefore for large $g$ and for not too small $\ep$
the errors of single and composite pulses scale as $\ep^2$ and
$\ep^4$, respectively. On the other hand, for $\ep/\gamma$ close to
1, errors of the composite pulses with larger $g$ are larger than
for smaller $g$. This is because the calculations in
Fig.~\ref{fig:gate_error1} refer to the same $\Delta/\eta^{1/2}$, in
which case the errors $\propto \ep^3,\ep^4$ are proportional to
$g^2$.

\section{Conclusions}

In this paper we have developed a theory of quantum gates based on
LZ pulses. In these pulses the control dc field is varied
in such a way that the qubit frequency passes through the frequency
of the external radiation field. In the adiabatic basis an LZ gate
can be expressed in a simple explicit form in terms of rotation
matrices. Our central result is that already a sequence of three LZ
pulses can be made fault-tolerant. The error of the corresponding
composite pulse ${\cal E}$ scales with the error $\ep$ in the qubit
energy or radiation frequency at least as $\ep^3$. In addition, the
coefficient at $\ep^3$ has an extra parametrically small factor. The
duration of the 3-pulse sequence is about 4 times the duration of
the single pulse, for the parameters that we used.

Fault tolerance of LZ gates is partly due to the change of state
populations being independent of precise frequency tuning. In
particular, LZ tunneling makes it possible to implement simple
$\pi$-pulses with an exponentially small error in the state
population.

The approach developed here can be easily generalized to more
realistic smooth pulses, as mentioned above. It can be applied also to
two-qubit gate operations in which the frequencies of interacting
qubits are swept past each other, leading to excitation transfer
\cite{Dykman2001b}.  Such operations are complementary to two-qubit
phase gates and require a different qubit-qubit interaction.

LZ pulses provide an alternative to control pulses where qubits stay
in resonance with radiation for a specified time
\cite{Vandersypen2004}. In this more conventional approach it is
often presumed that qubits are addressed individually by tuning
their frequencies. In contrast to this technique, LZ pulses do not
require stabilizing the frequency at a fixed value during the
operation. As a consequence, calibration of LZ pulses is also
different, which may be advantageous for some applications, in
particular in charge-based systems
\cite{Platzman1999,Smelyanskiy2005}. The explicit expressions
discussed above require that the qubit transition frequency vary
linearly with time, but the linearity is needed only for a short
time when the qubit and radiation frequencies are close to each
other, as seen from Fig.\ref{fig:crossing}, which should not be too
difficult to achieve.

For pulses based on resonant tuning for a fixed time, much effort
has been put into developing fault-tolerant pulse sequences, see
Ref.~\onlinecite{Brown2004} and papers cited therein. In particular,
for energy offset errors it has been shown that a three-pulse
sequence can reduce the error to ${\cal E}\sim\ep^2$
\cite{Cummins2003} (the fidelity $F$ evaluated in
Ref.~\onlinecite{Cummins2003} is related to ${\cal E}$ discussed in
Ref.~\onlinecite{Brown2004} and in this paper by the expression
$1-F\propto {\cal E}^2$ for small ${\cal E}$; therefore an error
${\cal E}\sim\ep^2$ corresponds to the estimate \cite{Cummins2003}
$1-F\sim\ep^4$). This error is parametrically larger, for small
$\ep$, than the error of the three-pulse sequence proposed here,
${\cal E}\propto\ep^3$. We note that, with two correcting pulses of
a more complicated form, it is possible to eliminate errors of
higher order in $\ep$.

It follows from the results of this paper that fault-tolerant LZ
gates can be implemented using the standard repertoire of control
techniques and may provide a viable alternative to the conventional
single qubit gates.

\begin{acknowledgments}
This work was supported in part by the NSF through grant ITR-0085922
and by the Institute for Quantum Sciences at Michigan State
University.
\end{acknowledgments}


\end{document}